# Polymer coated cerium oxide nanoparticles as oxidoreductase-like catalysts

Victor Baldim[a†], Yadav Nisha[b†], Nicolas Bia[c], Alain Graillot[c], Cédric Loubat[c], Sanjay Singh[b], Ajay S. Karakoti[d] and Jean-François Berret[a‡]

[a]Université de Paris, CNRS, Matière et systèmes complexes, 75013 Paris, France
[b]Division of Biological and Life Sciences, School of Arts and Sciences, Ahmedabad University, Navrangpura, Ahmedabad 380009, Gujarat, India
[c]SPECIFIC POLYMERS, ZAC Via Domitia, 150 Avenue des Cocardières, 34160 Castries, France
[d]Global Innovative Center for Advanced Nanomaterials (GICAN), Faculty of Engineering and Built Environment (FEBE), The University of Newcastle, Callaghan, NSW, Australia

**Abstract:** Cerium oxide nanoparticles have been shown to mimic oxidoreductase enzymes by catalyzing the decomposition of organic substrates and reactive oxygen species. This mimicry can be found in superoxide radicals and hydrogen peroxides, harmful molecules produced in oxidative stress associated diseases. Despite the fact that nanoparticle functionalization is mandatory in the context of nanomedicine, the influence of polymer coatings on their enzyme-like catalytic activity is poorly understood. In this work, six polymer coated cerium oxide nanoparticles are prepared by association of 7.8 nm cerium oxide cores with two poly(sodium acrylate) and four poly(ethylene glycol) (PEG) grafted copolymers with different terminal or anchoring end groups, such as phosphonic acids. The superoxide dismutase-, catalase-, peroxidase- and oxidase-like catalytic activities of the coated nanoparticles were systematically studied. It is shown that the polymer coatings do not affect the superoxide dismutase-like, impair the catalase-like and oxidase-like and surprisingly improves peroxidase-like catalytic activities of cerium oxide nanoparticles. It is also demonstrated that the particles coated with the PEG-grafted copolymers perform better than the poly(acrylic acid) coated ones as oxidoreductase-like enzymes, a result that confirms the benefit of having phosphonic acids as anchoring groups at the particle surface.

† These authors contributed equally to this work
‡ Corresponding author: **jean-francois.berret@univ-paris-diderot.fr**

# I - Introduction

Cerium oxide nanoparticles (CNPs) are non-stoichiometric $CeO_2$ nanocrystals with a cubic fluorite-like structure made of divalent oxygen anions $O^{2-}$, tetravalent $Ce^{4+}$ and surface clusters of trivalent $Ce^{3+}$ neighboring oxygen vacancies.[1-3] Cerium oxide nanoparticles have been used as heterogeneous catalysts on electron transfer reactions. In general, the catalytic processes consist in the adsorption of reactants at the particle surface, followed by the surface reaction of the adsorbed species then by desorption of the products.[4-5] Recently, it has been suggested that CNPs may also be suitable for biomedical applications by mimicking oxidoreductase in the decomposition of some endogenous organic molecules and reactive oxygen or nitrogen species. The species most often encountered in these reactions are superoxide radical anions ($O_2^{-\bullet}$), hydrogen peroxide ($H_2O_2$), hydroxyl radicals ($^\bullet OH$), nitric oxide (NO) and peroxynitrite ($ONOO^-$).[6-13] The nanomedicine applications of cerium oxide stems from the remarkable properties shown by the nanocrystals, such as sizes in the nanometer range, which allows the addition of surface functional groups and redox switching between $Ce^{3+}$ and $Ce^{4+}$ states, which confer antioxidant capacities to CNPs.[14-15] These properties make cerium oxide nanoparticles a promising candidate against oxidative stress related disorders, including cardiomyopathy, sepsis, multiple sclerosis and cerebral ischemic stroke.[16-22]

However, the use of CNPs for biomedical applications is still limited.[23] First the tendency to form aggregates in biological fluids reduces their specific surface area, and consequently their enzymatic activities. Second, interaction with proteins results in a corona formation, which further mitigates the particle performances. In the context of nanomedicine, it is hence important to synthesize CNPs that are both biocompatible and redox active. In practice, the long-term colloidal stability of nanoparticles in biological fluids is tailored by surface adsorbed or chemically bound water soluble macromolecules, either natural or synthetic, like DNA, polysaccharides, albumin, poly(acrylic acid), poly(ethylene glycol), poly(vinylpyrrolidone) among others.[24] Coating of biodegradable polymers not only provide the stability and biocompatibility, yet in some cases it was found to enhance the enzyme mimetic activities.[25] Likewise, coatings can also interfere with the enzymatic activity by decreasing the total number of available surface sites or increasing the diffusion length of the species to reach the surface active sites. Coatings can also impart entirely new enzymatic properties due to synergistic interaction with the surface or to the chemistry of the tethered polymers. Anchoring groups or ligands on the surface of the metal oxide nanoparticles hence plays a vital role in the definition of improved or reduced enzymatic performances.[26] On the other hand, they give a flexibility of obtaining different surface charges to the metal oxide surfaces varying from positive to negative zeta potentials that has been shown to influence the cellular internalization.

In this work, we use 7.8 nm cerium oxide nanoparticles coated with poly (acrylic acid) or poly(ethylene glycol) (PEG) copolymers as enzyme-mimicking heterogeneous catalysts for the decomposition of reactive oxygen species. These polymers were considered because they were found to prevent the adsorption of proteins and confer to metal oxide nanoparticles remarkable colloidal stability in biological media.[27-29] Keeping the core of the particles identical, and changing the polymers on their surface, the effect of polymer coats on the catalytic activities can be highlighted. In view of this goal, we examine the catalase-like, superoxide dismutase-like, peroxidase-like and oxidase-like catalytic activities of cerium oxide cores coated with either poly (acrylic acid) or PEG copolymers with different terminal or anchoring groups. We observe that the poly(ethylene glycol) coatings do not affect the superoxide dismutase-like, slightly impair the catalase-like and oxidase-like and surprisingly improves peroxidase-like catalytic activities of cerium oxide nanoparticles.

# II - Results and discussion

## II.1 - Polymer coated cerium oxide nanoparticles

$CeO_2$ nanoparticles were synthesized by thermo-hydrolysis of $Ce(NO_3)_4$.[30-31] Figure 1 shows a representative TEM micrograph of the bare particles (Figure 1a) together with the size distribution adjusted with a log-normal function and leading to a median diameter of 7.8 nm (Figure 1b). The hydrodynamic diameter of the bare particles was determined by dynamic light scattering at $D_H$ = 9.0 nm. Wide-angle X-ray scattering results **(Supplementary Information S1)** confirms the fluorite-like structure of the $CeO_2$ nanocrystals. XPS of Ce 3d electrons were used to quantify the fraction of $Ce^{3+}$ surface cations by dividing the area under the decomposed spectrum peaks coming from this species by the total area obtained from summation of all peaks, leading to 14.5% **(Supplementary Information S2)**. In the present work we seek to evaluate the influence of polymer coatings on the oxidoreductase-like catalytic activity of CNPs, keeping the core with constant morphological and physico-chemical properties.[32]

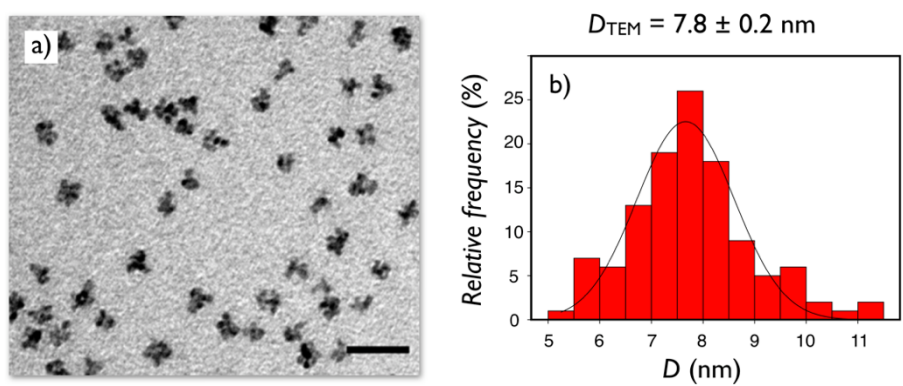

*Figure 1:*. *a) Transmission electron microscopy micrographs of bare cerium oxide nanoparticles. The scale bar is 25 nm. b) Particle size distributions adjusted to a log-normal function with median diameter $D_{TEM}$ = 7.8 nm and dispersity (ratio between the standard deviation and the average diameter) 0.15.*

An ensemble of six polymers was considered for coating the 7.8 nm CNPs (Figure 2). There are two linear poly(sodium acrylate) of molecular weight 2100 ($PAA_{2K}$) and 5100 g mol$^{-1}$ ($PAA_{5K}$), containing an average of 22 and 54 carboxylic acids per chain, respectively. The four other polymers are statistical copolymers synthesized by free radical polymerization. They contain phosphonic acid groups for surface attachment and poly(ethylene glycol) (PEG) chains for stability and protein resistance.[27,29] Phosphonic acid was considered in the syntheses because it is known to have a strong affinity towards metals or metal oxides (including cerium[33]) compared to sulfates and carboxylates, and it is anticipated that these residues build stronger links with the surface.[29,34-]

[38] The synthesis and characterization of the copolymers are described in **Supplementary Information S3 and S4**. Two of them are statistical polymers synthesized from two methacrylate comonomers, one bearing phosphonic acid groups (MPh), the other one bearing methyl terminated poly(ethylene glycol) chains (MPEG). The relative MPEG:MPh molar ratio in the copolymers is 0.50:0.50. The PEGs have molecular weight 2000 and 5000 g mol$^{-1}$, and the copolymers are named herein as MPEG$_{2K}$-MPh and MPEG$_{5K}$-MPh, respectively. The two remaining polymers are terpolymers containing in addition to the phosphonic acid groups, methyl terminated PEG of molecular weight 2000 g mol$^{-1}$ and amine terminated PEG chains of molecular weight 1000 or 2000 g mol$^{-1}$. The proportions of methyl terminated PEG, amine terminated PEG and phosphonic acids repeating units are 0.37:0.13:0.50 for MPEG$_{2K}$-MPEGa$_{1K}$-MPh and 0.25:0.25:0.50 for the MPEG$_{2K}$-MPEGa$_{2K}$-MPh, (where "a" stands for the amine terminal group). Poly (acrylic acid) homopolymers and phosphonic acid PEG copolymers have been shown to give excellent colloidal stability to metal oxide cores when these particles are dispersed in phosphate saline buffer or protein enriched culture medium.[27,29,39-40] The copolymers MPEG$_{2K}$-MPh, MPEG$_{5K}$-MPh, MPEG$_{2K}$-MPEGa$_{1K}$-MPh and MPEG$_{2K}$-MPEGa$_{2K}$-MPh were studied using static light scattering and found to have molecular weights $M_w^{Pol}$ of 20300, 22000, 39488 and 29200 g mol$^{-1}$ respectively. Assuming a molar mass dispersity Đ = 1.8[29], the number-averaged molecular weight $M_n^{Pol}$ was determined at 11300, 12200, 21900 and 16200 g mol$^{-1}$. From these values, the average number of phosphonic acid was estimated at 5.1, 2.3, 7.7 and 6.7. These later results are summarized in Table 1.

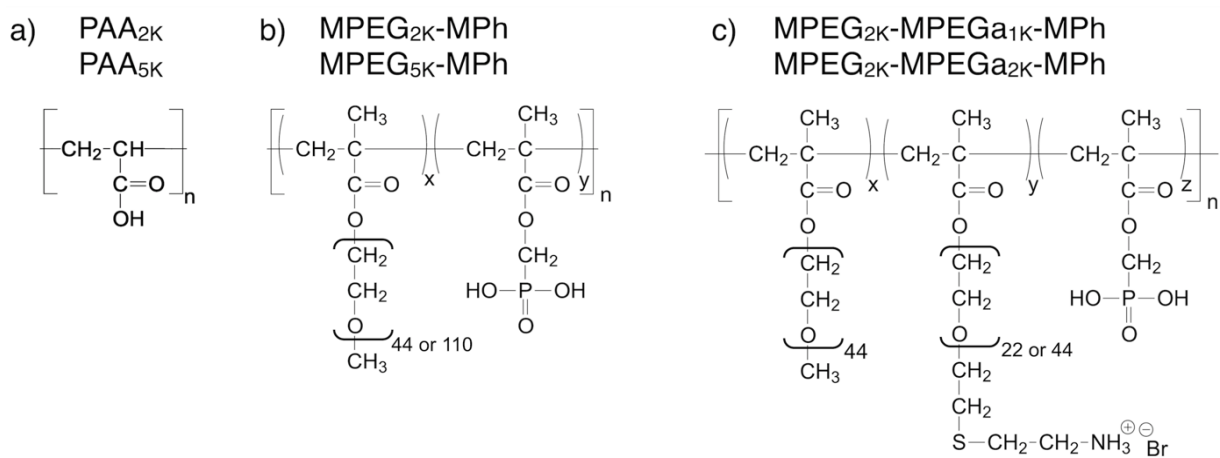

*Figure 2: Macromolecular structures of the polymers used for coating cerium oxide nanoparticles in this work. **a)** Poly(sodium acrylate) of molecular weight 2100 g mol$^{-1}$ (PAA$_{2K}$) or 5100 g mol$^{-1}$ (PAA$_{5K}$). **b)** Poly(poly(ethylene glycol)methacrylate-co-dimethyl(methacryoyloxy)methyl phosphonic acids) is a statistical copolymer synthesized from two methacrylate monomers, one bearing phosphonic acid groups (MPh), the other one bearing methyl terminated poly(ethylene glycol) (PEG) chains (MPEG) characterized by PEG chains of 2000 g mol$^{-1}$ (MPEG$_{2K}$-MPh) or 5000 g mol$^{-1}$ (MPEG$_{5K}$-MPh). **c)** A terpolymer analogous to MPEG$_{2K}$-MPh, however, containing also amine terminated PEG chains of 1000 g mol$^{-1}$ (MPEG$_{2K}$-MPEGa$_{1K}$-MPh) or 2000 g mol$^{-1}$ (MPEG$_{2K}$-MPEGa$_{2K}$-MPh).*

*Table 1*: *Molecular characteristics of the polymers and copolymers used as coatings for the cerium oxide nanoparticles shown in Figure 1 (**Supplementary Information S4**).*

| Polymers | $M_w^{Pol}$ (g mol$^{-1}$) | $M_n^{Pol}$ (g mol$^{-1}$) | Comonomer proportion (x:y:z) | Anchoring groups per polymer |
|---|---|---|---|---|
| PAA$_{2K}$ | 2100 | - | - | 22.3 |
| PAA$_{5K}$ | 5100 | - | - | 54.2 |
| MPEG$_{2K}$-MPh | 20280 | 11300 | 0.50:0.50:0 | 5.1 |
| MPEG$_{5K}$-MPh | 21960 | 12200 | 0.50:0.50:0 | 2.3 |
| MPEG$_{2K}$-MPEGa$_{1K}$-MPh | 39490 | 21900 | 0.37:0.13:0.50 | 7.7 |
| MPEG$_{2K}$-MPEGa$_{2K}$-MPh | 29160 | 16200 | 0.25:0.25:0.50 | 6.6 |

The cerium oxide nanoparticles were coated following the procedures described in the Materials and Methods section. In the following, the polymer coated nanoparticles are dubbed CeO$_2$@PAA$_{2K}$, CeO$_2$@PAA$_{5K}$, CeO$_2$@MPEG$_{2K}$-MPh, CeO$_2$@MPEG$_{5K}$-MPh, CeO$_2$@MPEG$_{2K}$-MPEGa$_{1K}$-MPh and CeO$_2$@MPEG$_{2K}$-MPEGa$_{2K}$-MPh. Upon mixing polymer and CNP dispersions, the polymers adsorb spontaneously at the particle surface through the carboxylic acids of the PAA$_{2,5K}$, or the phosphonic acids of the copolymers, leading to a core-shell structure. Carboxylic acids bind electrostatically to positively charged metal oxide surfaces and have been shown to bind strongly to cerium oxide surfaces.[41-43] Phosphonic acids bind metal oxides though condensation of their acidic hydroxyls P-OH with surface metal hydroxyls M-OH and/or coordination of the phosphoryl oxygen to Lewis acid surface sites.[44-46] The hydrodynamic diameters ($D_H$) and zeta potential ($\zeta$) of the coated CNPs were determined at physiological pH by dynamic light scattering and electrophoretic mobility measurements, respectively. These particles can be divided in three groups according to their electrostatic surface charge: the CNPs decorated with a poly(acrylic acid) shell are negatively charged,[41] those functionalized with PEGs are neutral[29] and those associated with an amine terminated PEGs are positively charged (Figure 3).

TEM micrographs of CeO$_2$@PAA$_{2,5K}$, CeO$_2$@MPEG$_{2,5K}$-MPh, and CeO$_2$@MPEG$_{2K}$-MPEGa$_{1,2K}$-MPh show individual particles separated from each other (**Supplementary Information Figure S5**). Average diameters for the particles were found to be 7.6 ± 1.1, 7.5 ± 1.1, 7.3 ± 0.8, 7.7 ± 1.1, 7.4 ± 0.9 and 7.5 ± 1.1 nm, respectively. These values are in agreement with that of the bare CNPs (Figure 1), suggesting that polymer coatings impart no noticeable change in particle size or morphology. With TEM, the polymer shell around the particles is not observed because of its low electronic contrast. For CeO$_2$@PAA$_{5K}$, some agglomerates are observed, which could be due to the formation of particle dimers and trimers. These agglomerates result from the fact that each PAA$_{5K}$ chain has on average 54 anchoring groups that are capable to bind to more than one particle.[41] The $D_H$'s of the coated CNPs, CeO2@PAA$_{2,5K}$, CeO$_2$@MPEG$_{2,5K}$-MPh and CeO$_2$@MPEG$_{2K}$-MPEGa$_{1,2K}$-MPh were found to be 15.1, 40.2, 27.2, 31.7, 29.5 and 31.5 nm, respectively (Table 2). When the polymer surface density is large (*i.e.* > 0.1 nm$^{-2}$),[47-48] the chains adopt a brush-type configuration, in which the chains are stretched radially and form a shell, as illustrated in Figure 3. This extended brush accounts for the increase of the hydrodynamic diameters, and its thickness ($h$) can be estimated. For $PAA_{2K}$, $h$ is equal to 3 nm whereas it is around 9 – 11 nm for the PEGs (Table 2).

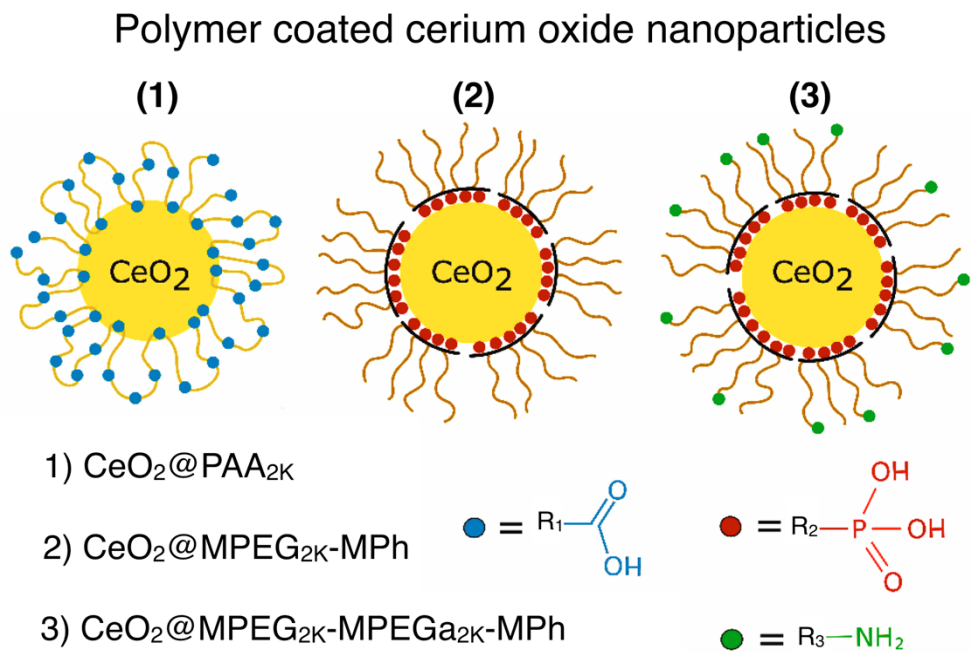

*Figure 3:* Representation of polymer coated cerium oxide nanoparticles investigated herein. Particle *1* is a $CeO_2@PAA_{2K}$ where acrylic acid repeating units depicted as blue spheres are anchored at CNP surface through carboxylic acid functions. Particle *2* is a $CeO_2@MPEG_{2K}$-MPh where several units of $MPEG_{2K}$-MPh are bound to the surface through phosphonic acid functions (red spheres). PEG chains of 2000 g mol$^{-1}$ grafted to the methacrylic backbone form a polymer brush. Particle *3* is a $CeO_2@MPEG_{2K}$-$MPEGa_{2K}$-MPh, where several units of $MPEG_{2K}$-MPh are bound to CNP surface as in the case of particle 2. The difference between them lies in the fact that some of the PEG chains are terminated by primary amines (green spheres).

*Table 2*: Hydrodynamic diameter ($D_H$), polymer brush thickness (h), zeta potential ($\zeta$) determined for polymer coated nanoparticles. The two last columns are the number of polymer per particle and the PEG density as determined from complementary measurements.[27,41] For $CeO_2@PAA_{5K}$ the large $D_H$ value is attributed to CNP aggregation, as identified by TEM.

| Nanoparticles | $D_H$ (nm) | $h$ (nm) | $\zeta$ (mV) | Polymers per particle | PEG density (nm$^{-2}$) |
|---|---|---|---|---|---|
| Bare CeO$_2$ | 9.0 | 0 | +21 ± 1 | - | - |
| CeO$_2$@PAA$_{2K}$ | 15.1 | 3.1 | -14 ± 2 | 45 | 0.24 |
| CeO$_2$@PAA$_{5K}$ | 40.2 | - | -15 ± 3 | - | - |
| CeO$_2$@MPEG$_{2K}$-MPh | 27.2 | 9.1 | +1.4 ± 0.1 | 15 | 0.28 |
| CeO$_2$@MPEG$_{5K}$-MPh | 31.7 | 11.4 | +0.6 ± 0.2 | 23 | 0.20 |
| CeO$_2$@MPEG$_{2K}$-MPEGa$_{1K}$-MPh | 29.5 | 10.2 | -1.1 ± 0.1 | - | - |
| CeO$_2$@MPEG$_{2K}$-MPEGa$_{2K}$-MPh | 31.5 | 11.2 | +5.8 ± 0.3 | 26 | 0.62 |

In parallel, electrophoretic mobility measurements show that bare CNPs are positively charged ($\zeta$ = +21 mV), CNPs coated with PAA$_{2,5K}$ are negatively charged ($\zeta$ = -15 mV)[41] and those coated with PEG copolymers neutral ($\zeta \sim 0$ mV). For the latter, charge neutrality results most likely from the neutralization of positive charges by the phosphonic acid groups after condensation of metal hydroxyl groups. For CeO$_2$@MPEG$_{2K}$-MPEGa$_{2K}$-MPh, the amine groups located in the outer brush region impart a slightly positive charge ($\zeta$ = +6 mV) to the particles. For CeO$_2$@MPEG$_{2K}$-MPEGa$_{1K}$-MPh in contrast, the amines are embedded within the brush, and the particle is again not charged (Table 2).

**II.2 - Catalase-like catalytic activity of cerium oxide nanoparticles**

The catalase-like catalytic activity of polymer coated CNPs was investigated by spectro-fluorimetry using the Amplex-Red reagent assay.[49-51] CNP dispersions at molar concentration $[CNP]$ = 0.1 – 300 nM were incubated with 5 µM hydrogen peroxide in 96-well plates. A mixture of horseradish peroxidase (HRP) and Amplex-Red was added to each well, this latter being transformed into the fluorescent resorufin due to the reaction with the remaining H$_2$O$_2$. The CAT-like activity is here defined as the percentage of decomposed H$_2$O$_2$ at the end of the assay. Figure 4a shows the CAT-like activity of bare and polymer coated particles, including CeO$_2$@PAA$_{2,5K}$, CeO$_2$@MPEG$_{2,5K}$-MPh and CeO$_2$@MPEG$_{2K}$-MPEGa$_{1,2K}$-MPh. Increasing the nanoparticle molar concentration leads to an increase in H$_2$O$_2$ disproportionation for the seven samples considered. For all, $A_{CAT}([CNP])$ follows a sigmoidal-shape dependence when plotted in a semi-logarithmic scale fitted using a Langmuir adsorption function of the form:[32]

$$A_{CAT}([CNP]) = \frac{1}{1 + [CNP]_0^{CAT}/[CNP]} \quad (1)$$

where $[CNP]_0^{CAT}$ is the only adjustable parameter. $[CNP]_0^{CAT}$ characterizes the CAT-like catalytic activity of a given particle in the conditions used. It also represents the concentration at which $A_{CAT}$ reaches half of the saturation value. To compare the effects of the coating on the CAT-like activity, we introduce two quantities, the turnover frequency $v_{TO}$ and the relative performance $P_{Rel}$. The turnover frequency is defined as $v_{TO} = [H_2O_2]/t[CNP]_0^{CAT}$ and denotes the number of hydrogen peroxide molecules decomposed per second and per nanoparticle, whereas the relative performance represents the ratio of the turnover frequency of a given particle with respect to that of the bare nanocrystals. Data show that bare CNPs are the most efficient CAT-like catalyst among the studied particles, with a value of $[CNP]_0^{CAT}$ of 12 nM and a $v_{TO}$ of 0.231 s$^{-1}$. For the PEGylated CNPs, similar sigmoidal behaviors are observed, leading to $[CNP]_0^{CAT}$ = 29 nM, $v_{TO}$ = 0.096 s$^{-1}$ and $P_{Rel}$ = 41%. Previous work has shown that the CAT-like catalytic activity of cerium oxide nanoparticles is correlated to the surface concentration of Ce$^{3+}$ but also to the amount of hydrogen peroxide adsorbed at the interface.[32] The previous results suggest the phosphonic acid bound to hydroxyl groups prevent H$_2$O$_2$ molecules or reactive intermediates from adsorbing onto it, decreasing the number of collisions leading to product formation. Interestingly, the catalytic activity is not affected by the copolymer structure. For the PAA$_{2,5K}$ coated CNPs, the $A_{CAT}$ concentration dependences are different, $[CNP]_0^{CAT}$ being equal to 600 and 130 nM. These coating are less efficient, the relative performances $P_{Rel}$ being equal to 2% and 9%. The differences between the two last samples could come from the slight aggregation seen in CeO$_2$@PAA$_{5K}$ dispersions or to the rather tight binding of carboxylic acid groups with both Ce$^{3+}$ and Ce$^{4+}$ surface sites.[42-43] These data are summarized in Table 3.

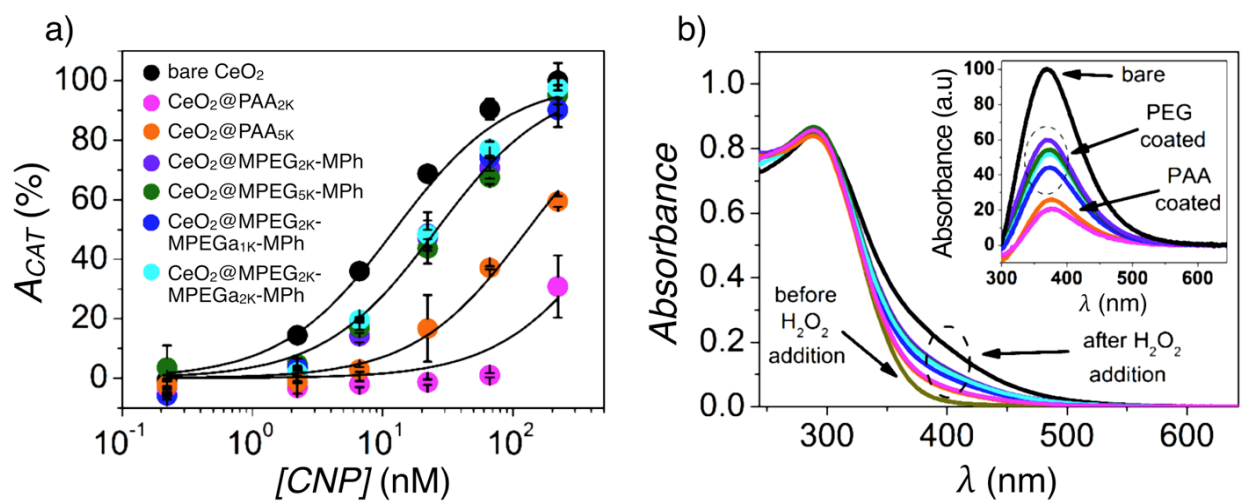

*Figure 4*: *a) Catalase-like catalytic activity of cerium oxide nanoparticles. Percentages of disproportionated hydrogen peroxide A$_{CAT}$ obtained for bare and the polymer coated CeO$_2$ nanoparticles. The black lines are the best fit of a single parameter Langmuir–like function (Eq.1). In this assay, the H$_2$O$_2$ concentration was initially set at 5 μM. b) UV-vis spectrophotometric measurement of the absorbance of CNPs dispersions in Tris buffer pH 7.5 before and after addition of H$_2$O$_2$. The initial and final CNP concentrations are 44.3 nM and 16.3 nM respectively. Inset: redshift absorption peak around 368 nm obtained after subtracting the absorbance from dispersions devoid of H$_2$O$_2$.*

*Table 3*: *CAT-like catalytic parameters for the bare and polymer coated cerium oxide nanoparticles.*

| Nanoparticles | $[CNP]_0^{CAT}$ (nM) | $\nu_{TO}$ (s$^{-1}$) | $P_{Rel}$ (%) |
|---|---|---|---|
| CeO$_2$ bare | 12 | 0.23 | 100 |
| CeO$_2$@PAA$_{2K}$ | 600 | 0.005 | 2 |
| CeO$_2$@PAA$_{5K}$ | 130 | 0.021 | 9 |
| CeO$_2$@MPEG$_{2K}$-MPh | 29 | 0.096 | 41 |
| CeO$_2$@MPEG$_{5K}$-MPh | 29 | 0.096 | 41 |
| CeO$_2$@MPEG$_{2K}$-MPEGa$_{1K}$-MPh | 29 | 0.096 | 41 |
| CeO$_2$@MPEG$_{2K}$-MPEGa$_{2K}$-MPh | 29 | 0.096 | 41 |

Previous work using UV-Vis measurements has shown that CNPs with CAT-like catalytic activity present a characteristic absorption peak upon $H_2O_2$ addition.[32,52-53] This peak appears as a red-shifted shoulder and vanishes over time as the disproportionation reaction proceeds. UV-Vis absorbance assays were performed on bare $CeO_2$, $CeO_2$@$PAA_{2,5K}$, $CeO_2$@$MPEG_{2,5K}$-MPh and $CeO_2$@$MPEG_{2K}$-$MPEGa_{1,2K}$-MPh (Figure 4b). The continuous curve labelled "before $H_2O_2$ addition" displays the absorption spectra obtained for the seven particles investigated, which are all superimposed. Upon $H_2O_2$ addition, the absorption spectra display a systematic red shift that is dependent on the coating. In the inset, absorption peaks centered around 368 nm were obtained after subtraction of the "before $H_2O_2$ addition" background spectrum. This figure shows that the effect of the polymer coating can also be assessed by examining the magnitude of the red shift. Concerning the interpretation of the above phenomenon, it was recently suggested that $H_2O_2$ molecules reacts with cerium oxide nanoparticles to form a stable peroxo and/or hydroperoxo species at the particle surface.[54-57] We propose here that the coordination sites for peroxide species are the critical factors in the CNP anti-oxidation process with $H_2O_2$. In this regard, coordinated peroxide species might be responsible for the $H_2O_2$ induced color changes. The UV-Vis spectra from Figure 4b confirm this tendency: the polymer coating hinders hydrogen peroxide from adsorption and decomposition, a fundamental step in the CAT-like catalytic activity of these nanoparticles.

**II.3 - Superoxide dismutase-like catalytic activity**

The SOD-like catalytic activity of CNPs was investigated by a colorimetric assay using UV-Vis spectroscopy.[49,58] With this assay, superoxide radical anions are generated *in situ* by xanthine oxidase. A water-soluble tetrazolium salt is then added to the solution, this latter being oxidized by the remaining superoxide radicals into a soluble formazan dye. The final product is then identified from its absorption property.[49-50,59] The SOD-like activity is defined as the percentage of dismutated superoxide radicals at the end of the reaction. Figure 5 shows the concentration dependence of $A_{SOD}$ in the range $10^{-2} - 10^3$ nM for the seven particles studied. It is found that the SOD-like activities of the bare and coated CNPs are well superimposed and follow a sigmoid-type behavior (continuous curve in Figure 5). The data were adjusted using a Langmuir-type adsorption isotherm similar to that of Eq. 1:[32]

$$A_{SOD}([CNP]) = \frac{1}{1 + [CNP]_0^{SOD}/[CNP]} \quad (2)$$

where $[CNP]_0^{SOD}$ is the only adjustable parameter. For the 7 samples, we found $[CNP]_0^{SOD} = 20.9$ nM, suggesting that the SOD-like catalytic activity does not depend on the coating or on the molecular structure of the tethered chains. In terms of relative performance, all coated particles have a $P_{Rel}$ of the order of 1. It is interesting to note that while catalase activity is affected by the polymers, the SOD activity has been relatively unperturbed by the coatings. Possible reasons for these observations are the following: *i)* the diffusion and adsorption of superoxide on the particle surface is unaffected by the coatings; *ii)* as the SOD activity is largely dependent on the $Ce^{3+}$ concentration, these coatings do not bond tightly to $Ce^{3+}$ surface, therefore superoxide radicals can compete for $Ce^{3+}$ surface more actively; *iii)* the $Ce^{3+}$ sites are unaffected by surface coatings and the ratio of $Ce^{3+}/Ce^{4+}$ do not change significantly; *iv)* the coatings anchor on the surface of cerium oxide nanoparticles through metal centers while the superoxide may adsorb on the oxygen defect sites and participate in the redox reaction. One or all of these reasons together contribute to this unique SOD activity of CNPs.

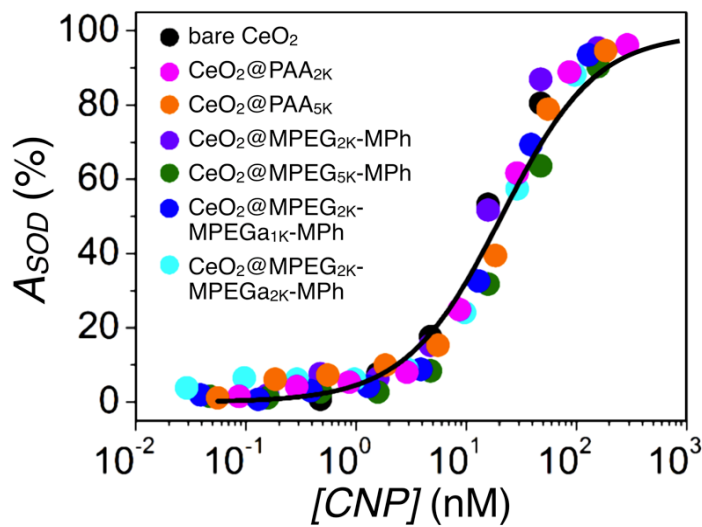

***Figure 5:*** *Percentage of dismutated superoxide radicals $A_{SOD}$ obtained for bare $CeO_2$, $CeO_2@PAA_{2,5K}$, $CeO_2@MPEG_{2,5K}$-MPh and $CeO_2@MPEG_{2K}$-MPEg$_{1,2K}$-MPh. The black line results from least square fitting using Eq. 2, the adjustable parameter $[CNP]_0^{SOD}$ being 20.9* nM.

**II.4 – Auto-regeneration ability of cerium oxide nanoparticles**

Auto-regeneration is a unique property of CNPs by which the particles can switch between the two oxidation states, $Ce^{3+}$ and $Ce^{4+}$ and regenerate catalytic surface activity. To study the effect of polymer coating on the auto-regeneration, 0.2 M $H_2O_2$ was added to 1 g L$^{-1}$ bare and coated CNPs (Figure 6a). Tubes 1 to 7 in the subset images represent $CeO_2$, $CeO_2@MPEG_{2,5K}$-MPh, $CeO_2@MPEG_{2K}$-MPEg$_{1,2K}$-MPh and $CeO_2@PAA_{2,5K}$ respectively. The addition of $H_2O_2$ almost immediately converts the colorless dispersions to yellow colored dispersions (Figure 6b), indicating the oxidation of $Ce^{3+}$ state to $Ce^{4+}$. Subsequently, the SOD-like activity was measured by following the reduction of cytochrome C as a function of the time, before and after addition of $H_2O_2$. Bare and polymer coated CNPs showed superoxide radical scavenging ability, as displayed by the decrease in absorbance at 550 nm (Figure 6a). Further, the oxidized CNPs, after $H_2O_2$ exposure, showed a comparatively less significant decrease in absorbance, which could be due to the conversion of $Ce^{3+}$ to $Ce^{4+}$ oxidation state by $H_2O_2$ (Figure 6b). The conversion of yellow dispersions to colorless after 15 days of incubation is an indication of the regeneration of the CNP surfaces, which was confirmed by the restoration of SOD activity in the regenerated samples (Figure 6c).

The slight differences observed in the absorbance results of Figures 6b and 6c suggest that the CNP samples did not lose their SOD activity at the same rate, a result that can be ascribed to their different catalase mimetic activity. Thus, different samples can decompose portions of the $H_2O_2$ added to oxidize $Ce^{3+}$ to $Ce^{4+}$ at a different rate resulting in slight changes in the regeneration of their surfaces.

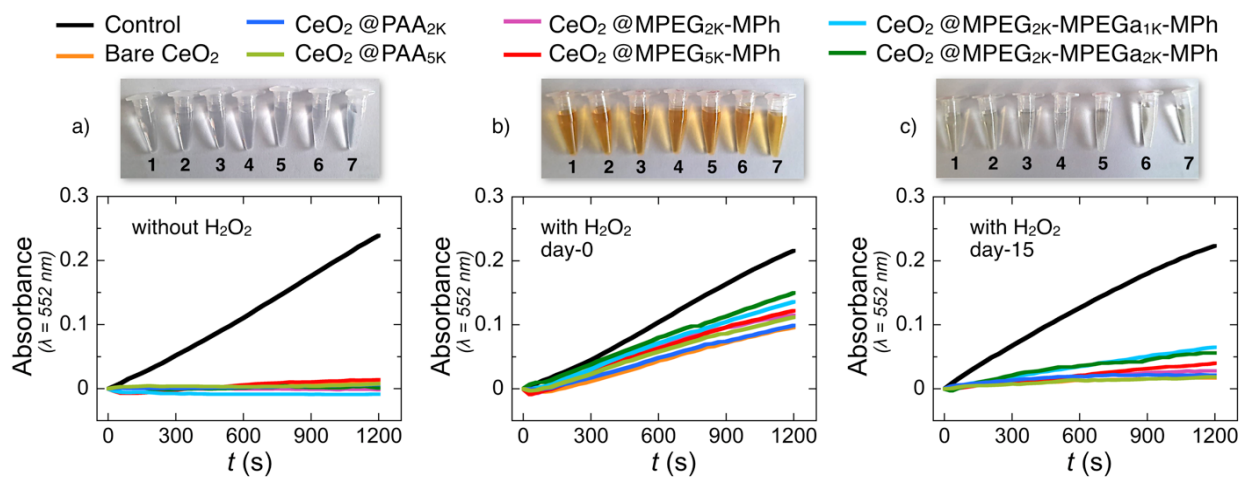

*Figure 6: Comparison of auto-regenerative ability of cerium oxide nanoparticles tested as a function of loss in SOD activity and its recovery before (a) immediately after (b) and after 15 days (c) of addition of $H_2O_2$. Before addition of $H_2O_2$, bare and coated particles show SOD activity ([CNP] = 100nM). Addition of $H_2O_2$ results in loss of SOD activity due to oxidation of surface $Ce^{3+}$ ions into $Ce^{4+}$ ions. After 15 days, the CNPs autoregernerate their surface and recover their SOD catalytic activity. Change in color from colorless to yellow following $H_2O_2$ addition is shown in subset images where samples 1 to 7 are bare $CeO_2$, $CeO_2@MPEG_{2,5K}$-MPh, $CeO_2@MPEG_{2K}$-$MPEGa_{1,2K}$-MPh and $CeO_2@PAA_{2,5K}$ respectively. After a 15-day incubation yellow color disappears in all samples. Absorbance of cytochrome C at 550 nm in absence of particles is used as control.*

### II.5 – Peroxidase-like catalytic activity of cerium oxide nanoparticles

The peroxidase-like catalytic activity of bare and coated CNPs was investigated by using the particles as heterogeneous nanocatalysts in the oxidation of the 3,3',5,5'-tetramethylbenzidine (TMB) by hydrogen peroxide. The concentration of the oxidized TMB can be determined by absorbance spectroscopy in the near-UV and visible ranges. In a first step, we investigated peroxidase-like activity in acetate buffer at pH 4.0. as a function of the coating (Figure 7a) and time (Figure 7b) by following the oxidation of the TMB substrate by $H_2O_2$. Here, the dispersions studied were $CeO_2@PAA_{2,5K}$, $CeO_2@MPEG_{2,5K}$-MPh at concentrations of 10 nM. The spectral variation of catalytic activity was identified from the appearance of strong absorption peaks at 370 and 652 nm, this later being associated with appearance of blue color (inset of Figure 7a). The enhanced peroxidase-like activity was observed for all coated CNPs with respect to bare particles or to the control. Note that for PEGylated CNPs the normalized absorbance grows continuously with time, whereas for the poly(acrylic acid) coating, it passes through a maximum at 4 min after the addition of $H_2O_2$ and decreases to a lower level (Figure 7b). This outcome suggests that PEGylated CNPs show single electron oxidation reaction, whereby the colorless TMB substrate (reduced form) is converted into greenish-blue product (partially oxidized form). CeO$_2$@PAA$_{2,5K}$ in contrast complete 2 electron oxidations of the TMB substrate, for which the color changes from colorless to blue during the first electron oxidation and then from blue to yellow during the second.

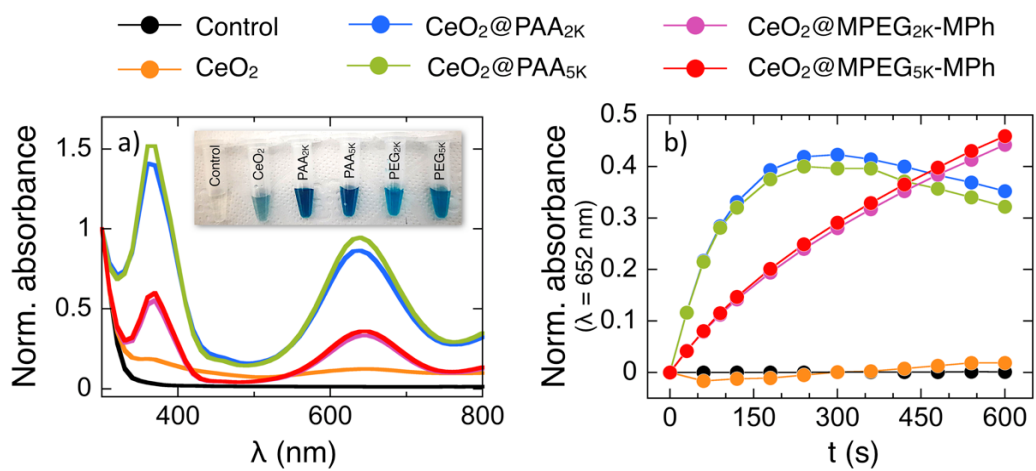

*Figure 7*: **a)** *Comparative spectral variation of peroxidase mimetic activity of TMB oxidation by hydrogen peroxide in acetate buffer 0.1 M (pH 4.0) catalyzed by bare and polymer coated CNPs. **Inset**: images of samples after CNP addition. **b)** Kinetics study of peroxidase-like activity of coated CNPs with different polymer coatings. In the two experiments, the CNP, TMB and H$_2$O$_2$ concentrations were fixed at 10 nM, 0.8 mM and 2 M, respectively.*

Subsequently, we focused on the reaction kinetics and on the effect of the CNP concentration. Figure 8a highlights the growth of the 370 and 652 nm absorbance peaks with time in the presence of 10 nM CeO$_2$@MPEG$_{2K}$-MPh particles. In this assay, the experiment lasted 600 s and the TMB and H$_2$O$_2$ concentrations were fixed at 0.8 mM and 2 M, respectively. The experiment was repeated varying [CNP] from 1 to 20 nM and the absorbance at 652 nm was recorded over time (Figure 8b). The data show that the peak amplitude at 652 nm increases continuously over 10 min after H$_2$O$_2$ addition. In a next assay, the concentrations of CeO$_2$@MPEG$_{2K}$-MPh and H$_2$O$_2$ were fixed at 10 nM and 2 M and that of TMB was varied from 0.05 to 0.8 mM, allowing to measure the time evolution of the oxidized form of the TMB substrates (Figure 8c). Focusing on the initial rate of product formation $v$, which is determined from the initial slope of the time dependent data, the peroxidase activity can be quantified. Figure 8d displays the evolution of the initial rate for CeO$_2$@MPEG$_{2K}$-MPh at 10 nM. The data exhibit a behavior characterized by a linear increase followed by a saturation plateau. They also reveal that the peroxidase catalytic activity of the polymer coated CNPs is enhanced compared to bare particles. The peroxidase-like activity of CNP was assessed using the Michalis-Menten steady state equation:

$$v([TMB]) = V_{Max} \frac{[TMB]}{K_M + [TMB]} \tag{3}$$

$$V_{Max} = K_{CAT}[CNP]_0 \tag{4}$$

In the previous equations, $[CNP]_0$ is the molar concentration of nanoparticles, $V_{Max}$ the maximum rate of product formation, $K_M$ the Michaelis-Menten constant or the amount of TMB required for the rate of product formation to be $V_{Max}/2$. In Eq. 4, $K_{CAT}$ denotes the amount of products formed per unit of time on a single nanoparticle, or the maximum frequency of product formation. The continuous lines in Figure 8d attest to the good agreement between the data and the model. The fitting parameters are listed in Table 4.

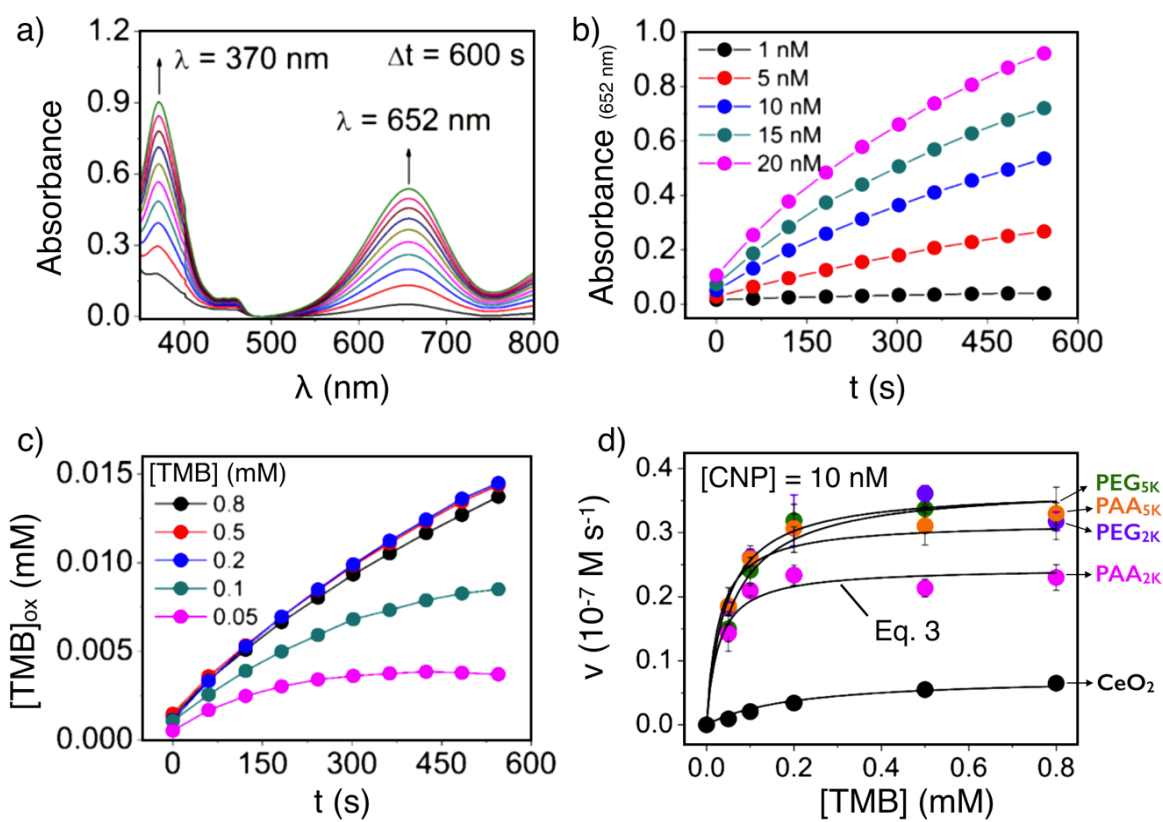

*Figure 8*: *a) Spectral variation of TMBox originated from the oxidation of TMB by hydrogen peroxide in 0.1 M acetate buffer (pH 4.0) at room temperature. The particle used here are CeO₂@MPEG₂ₖ-MPh at [CNP] = 10 nM. In this assay, [TMB] = 0.8 mM and [H₂O₂] = 2 M. The production of TMBox can be monitored by following its absorbance at 370 and 652 nm. b) Evolution of the absorption at 652 nm for CeO₂@MPEG₂ₖ-MPh varying the [CNP] from 1 - 20 nM with [TMB] = 0.8 mM and [H₂O₂] = 2 M. c) Time evolution for the concentration of oxidized TMB using [CeO₂@MPEG₂ₖ-MPh] =10 nM, [H₂O₂] = 2 M and varying [TMB] from 0.05-0.8 mM. d) Michaelis-Menten curves for the oxidation of TMB catalysed by bare and polymer coated cerium oxide nanoparticles CeO₂@PAA₂,₅ₖ and CeO₂@MPEG₂,₅ₖ-MPh at [CNP] = 10 nM. The continuous curves are from Eqs. 3 and 4.*

A remarkable feature from Figure 8 is that $K_M$ values are always smaller for TMB than for $H_2O_2$. $K_M$ (TMB) and $K_M$ ($H_2O_2$) obtained for horseradish peroxidase,[60] and for CeO₂@PAA₂ₖ, CeO₂@MPEG₂ₖ-MPh are $4.3\times10^{-4}$ and $3.7\times10^{-3}$ M, $4.3\times10^{-5}$ and $5.8\times10^{-2}$ M, $2.7\times10^{-5}$ and $1.6\times10^{-1}$ M respectively. The TMB concentration necessary for $v$ to be half of the maximum is more than 5000 smaller than that of $H_2O_2$ when using CeO₂@MPEG₂ₖ-MPh as catalyst. This means that the affinity of TMB for the active sites is larger than that of $H_2O_2$. Another remarkable feature is that $K_{CAT}$ values for TMB and $H_2O_2$ for a given nanoparticle are close. $K_{CAT}$ (TMB) and $K_{CAT}$ ($H_2O_2$) obtained for horseradish peroxide,[60] $CeO_2$, $CeO_2$@$PAA_{2K}$ and $CeO_2$@$MPEG_{2K}$-MPh are $4.0 \times 10^3$ and $3.5 \times 10^3$ s$^{-1}$, $7.5 \times 10^{-1}$ and $6.5 \times 10^{-1}$ s$^{-1}$, 3.7 and 3.0 s$^{-1}$, 2.45 and 2.2 s$^{-1}$ respectively. Even if the affinity of TMB to the active sites is higher than that of $H_2O_2$, the rate of the one electron oxidation of TMB is limited and, therefore, the same as the rate of one electron reduction of $H_2O_2$. The effect of polymer coatings in the peroxidase-like catalytic activity can be evaluated by means of their catalytic efficiency, defined as $K_{CAT}/K_M$. Table 4 shows that polymer coated particles have 20 times larger ratios $K_{CAT}/K_M$ as compared to bare CNPs. Similar enhanced activities of the peroxidase activity of magnetite and gold nanoparticles were recently reported.[61-63] Oxidase activity of bare and coated CNPs was also evaluated at concentration of 250 nM and revealed an overall decrease of the activity in presence of the polymers **(Supplementary Information S6)**.

*Table 4: TMB peroxidase enzyme-like parameters for cerium oxide nanoparticles dispersed in acetate buffer pH 4.0.*

| Nanoparticles | [CNP] (nM) | $K_M$ (M) | $K_{CAT}$ (s$^{-1}$) | $K_{CAT}/K_M$ (M$^{-1}$ s$^{-1}$) |
|---|---|---|---|---|
| $CeO_2$ bare | 10 | $2 \times 10^{-4}$ | 0.75 | $3.8 \times 10^3$ |
| $CeO_2$@$MPEG_{2K}$-MPh | 10 | $4.3 \times 10^{-5}$ | 3.68 | $8.5 \times 10^4$ |
| $CeO_2$@$MPEG_{5K}$-MPh | 10 | $4.0 \times 10^{-5}$ | 3.60 | $9.0 \times 10^4$ |
| $CeO_2$@$PAA_{2K}$ | 10 | $2.7 \times 10^{-5}$ | 2.45 | $9.0 \times 10^4$ |
| $CeO_2$@$PAA_{5K}$ | 10 | $3.9 \times 10^{-5}$ | 3.45 | $8.9 \times 10^4$ |

The TMB can be oxidized by superoxide or hydroxyl radicals and it is important to differentiate whether the peroxidase activity is due to the production of hydroxyl radicals •OH from $H_2O_2$ or from other ROS species. To check the hydroxyl radicals produced during the reaction, the fluorescence intensity of conversion of terephthalic acid to 2-hydroxyterephthalic acid (2HTA) was monitored **(Supplementary Information S7)**. Terephthalic acid is a non-fluorescent molecule which converts to 2HTA in the presence of •OHs.[64] Results showed that there was an increase in hydroxyl radical generation after polymer coating and that the maximum hydroxyl radicals were observed with $PAA_{2,5K}$, in agreement with peroxidase activity data (Figure 7). The •OH generation of $CeO_2$@$MPEG_{2K}$-MPh was further investigated in a concentration dependent manner. The fluorescence intensity of 2-HTA was found to increase with the concentration of $CeO_2$@$MPEG_{2K}$-MPh, suggesting an additional production of hydroxyl radical **(Supplementary Information S8)**. At 1 nM $CeO_2$@$MPEG_{2K}$-MPh, hydroxyl radicals were found to be 7.9 nanomoles, increasing to 23 nanomoles at 20 nM concentration. To further confirm that the variation in fluorescence intensity is ascribed to •OH radical production, scavenging effect of ethanol was finally studied. In presence of ethanol, the inhibition of hydroxyl radical generation was observed for all coated CNPs and at all concentrations of $CeO_2$@$MPEG_{2K}$-MPh. These results confirmed that hydroxyl radical generation is responsible for increased peroxidase-like activity.

# III - Conclusion

In this article, we report on cerium oxide nanoparticles structural and antioxidant properties, and on the synthesis of functional polymers used as coating. The polymers examined have a dual functionality, one for protection against protein adsorption and one for targeting with terminal amine groups allowing additional covalent binding. As a comparison, we also examine coatings made form poly(acrylic acid) of different molecular weights, these polymers imparting negative surface charges to the cerium oxide particles. For the two types of polymer coats, we show that the particles remain dispersed and not aggregated (except for $PAA_{5K}$ which exhibits the presence of dimers and trimers), with a core of 7.8 nm and a corona of 3 - 10 nm thick. Keeping the CNP core identical and changing the nature of the coat, we are able to specifically highlight the impact of polymers on the oxidoreductase catalytic activities of cerium oxide nanoparticles. The main result that emerges from this work is that functionalization with polymers can tune the enzyme mimetic activities of cerium oxide nanoparticles. In our experiments, it is found that polymers do not affect the superoxide dismutase-like, slightly impair the catalase-like and oxidase-like and surprisingly improves peroxidase-like catalytic activities of the particles. It is also evidenced that CNPs coated with the PEG-grafted copolymers perform better than the poly(acrylic acid) coated ones as oxidoreductase-like enzymes. The data suggest that the impact of polymers on the catalase and SOD activities is most likely due to the ability of polymers to interact with metal centers and modify their oxidation state, or even the ability of ROS to diffuse and access the active sites. On the other hand, the peroxidase activity of CNPs is likely to stem from the synergistic action between the polymers and the particles. As the selectivity of nanozymes is a critical issue in biomedical applications, these surface functionalization strategies present an opportunity to suppress or enhance their enzymatic activity, while providing excellent stability and dispersion in biofluids. The summary of the knowledge obtained by this study is that biocompatible polymers can be pursued as surface modifiers of cerium oxide nanoparticles that can tune its enzymatic activities.

# IV - Materials and methods

### IV.1 - Materials

Cerium oxide nanoparticles (CNP) with a nominal diameter of 7.8 nm were kindly provided by Solvay (Centre de Recherche d'Aubervilliers, Aubervilliers, France). The poly (poly(ethylene glycol) methacrylate-co-dimethyl-(methacryoyloxy)methyl phosphonic acid) polymers, bearing phosphonic acids and PEG chains were synthesized by Specific Polymers® (Castries, France) and are abbreviated as $MPEG_{2K}$-MPh, $MPEG_{5K}$-MPh, $MPEG_{2K}$-$MPEGa_{1K}$-MPh and $MPEG_{2K}$-$MPEGa_{2K}$-MPh.[27,29] Details of the synthesis and molecular characteristics can be found in **Supplementary Information S3** and in Table 1. Poly(acrylic acid) sodium salts of weight-averaged molecular weight $M_w$ = 2100 g mol$^{-1}$ ($PAA_{2K}$) and 5100 g mol$^{-1}$ ($PAA_{5K}$), hydrogen peroxide ($H_2O_2$, 30 vol. %) and 3,3',5,5'-tetramethylbenzidine (TMB) were obtained from Sigma-Aldrich (Lyon, France).

### IV.2 - Polymer coated nanoparticles

The poly(acrylic acid) coated nanoparticles (CeO$_2$@PAA$_{2K}$ and CeO$_2$@PAA$_{5K}$) were prepared according to the precipitation-redispersion protocol.[41,65] The particles coated with the phosphonic acid PEG copolymers are noted CeO$_2$@MPEG$_{2K}$-MPh and CeO$_2$@ MPEG$_{5K}$-MPh and those made with terpolymers bearing amine groups at the PEG ends CeO$_2$@MPEG$_{2K}$-MPEGa$_{1K}$-MPh and CeO$_2$@MPEG$_{2K}$-MPEGa$_{2K}$-MPh. The coated particles were prepared as follows:[27,29,33,40] a dispersion of CeO$_2$ nanoparticles was prepared at a 2 g L$^{-1}$ in HNO$_3$ (pH 1.5). Co- and terpolymer solutions were prepared to a concentration of 2 g L$^{-1}$ in HNO$_3$ (pH 1.5). The stock dispersion and stock polymer solutions were filtered with Millipore filter 0.22 μm. The CNP dispersion was added dropwise to the polymer solution under magnetic stirring keeping the mixing volume ratio at $X_C/5$ where $X_C$ denotes the critical mixing ratio (nanoparticle over polymer) above which the CNPs are partially coated and precipitate at physiological pH. Working at $X_C/5$ insures that the polymers are in excess during adsorption.[27,33] After increasing their pHs to 8 by addition of NH$_4$OH, the dispersions were centrifuged at 4000 rpm using Merck centrifuge filters (pore 100000 g mol$^{-1}$) to remove the polymer excess and further concentrated to 20 g L$^{-1}$.

### IV.3 - Static and Dynamic Light Scattering

The scattered intensity and the hydrodynamic diameter $D_H$ were obtained from the NanoZS Zetasizer spectrometer (Malvern Instruments) with detection angle at 173°. The second-order autocorrelation function was analysed using the cumulant and CONTIN algorithms to determine the average diffusion coefficient $D_C$ of the scatterers. $D_H$ was calculated according to the Stokes-Einstein relation $D_H = k_B T/3\pi\eta D_C$ where $k_B$ is the Boltzmann constant, $T$ the temperature and $\eta$ the solvent viscosity. The hydrodynamic diameters provided here are the second coefficients in the cumulant analysis, noted $Z_{Ave}$ in the Malvern software. Measurements were performed in triplicate at 25 °C after an equilibration time of 120 s.

### IV.4 - Electrophoretic mobility and zeta potential

Laser Doppler velocimetry using the phase analysis light scattering mode and detection at an angle of 16° was used to carry out the electrokinetic measurements of electrophoretic mobility and zeta potential with the Zetasizer Nano ZS equipment (Malvern Instruments, UK). Zeta potential was measured after a 120 s equilibration at 25 °C.

### IV.5 - Ultraviolet-visible Spectroscopy

A UV-visible spectrometer (SmartSpecPlus from BioRad) was used to measure the absorbance of polymer coated cerium oxide nanoparticles aqueous dispersions. Absorbance data were used to determine the nanoparticle concentration for each batch by means of Beer-Lambert law.

### IV.6 - Transmission Electron Microscopy (TEM)

Micrographs were taken with a Tecnai 12 TEM operating at 80 kV equipped with a 1K×1K Keen View camera. CNP dispersions were deposited on ultrathin carbon type-A 400 mesh copper grids (Ted Pella, Inc.). Micrographs were analyzed using ImageJ software for 200 particles. The surface $S$ of each particle was measured manually, and the diameter $D$ was calculated from the expression: $D = \sqrt{4S/\pi}$. The Ferret analysis performed by the same plugin also provides the maximum and minimum diameters $D_{Max}$ and $D_{Min}$, from which the diameter was again estimated: $D = 1/2(D_{Max} + D_{Min})$. For the cerium oxide nanoparticles investigated here, the two determinations give the same size distribution (Fig. 1b). The particle size distribution was adjusted using a lognormal function of the form $p(d, D, s) = \frac{1}{\sqrt{2\pi}\beta(s)d} exp\left(-\frac{ln^2(d/D)}{2\beta(s)^2}\right)$, where $D$ is the median diameter and $\beta(s)$ is related to the size dispersity $s$ through the relationship $\beta(s) = \sqrt{ln(1 + s^2)}$. $s$ is defined as the ratio between the standard deviation and the average diameter.[65-66] The TEM experiments and analysis were performed for the bare and for the coated particles, allowing to confirm that coated particles were not aggregated following the previous protocol.

### IV.7 - SOD mimetic activity assay

The catalytic activity of cerium oxide nanoparticles in the dismutation of superoxide radical anion was assessed by a colorimetric assay using UV-Vis spectroscopy (Kit #19160-1KTF).[49,58] Briefly, 20 µL of a CNP dispersion in Tris-Cl buffer pH 7.5 was added to a well of a 96-well plate and mixed with 200 µL of WST-1 (2-(4-Iodophenyl)-3-(4-nitrophenyl)-5-(2,4-disulfophenyl)-2H-tetrazolium, monosodium salt). The reaction was initiated with the addition of 20 µL of xanthine oxidase solution, prepared by mixing 5 µL of the enzyme in 2.5 mL of a dilution buffer provided. After incubating plate at 37°C for 20 min, the absorbance at 450 nm was measured using a microplate reader (EnSpire Multimode Plate Reader, Perkin Elmer). Final CNP dispersion concentration ranged from 200 to 2000 µM. The SOD-like activity, noted $A_{SOD}$ is defined as the percentage of dismutated superoxide radicals after the 20 min.

### IV.8 – Auto regeneration ability

The self-regeneration ability of CNPs (1 g L$^{-1}$) was investigated by addition of 0.2 M H$_2$O$_2$. Bare and polymer coated particles (1 g L$^{-1}$) were incubated with 0.2 M H$_2$O$_2$ which oxidized cerium oxide from Ce$^{3+}$ oxidation state to Ce$^{4+}$ oxidation state. Oxidation converts colorless solution into yellow color solution which indicates more Ce$^{4+}$ on the lattice particle surface. SOD-mimetic activity was followed before and after addition of H$_2$O$_2$ by using UV-Vis spectrophotometer (Biotek, Synergy HT spectrophotometer) by the method elucidated by Korsvik *et al.*[67] To check the SOD mimetic activity, ferri-cytochrome C reduction was observed by using UV-Vis spectrophotometer (Biotek, Synergy HT spectrophotometer). In brief, 100 nM of cerium oxide nanoparticle dispersion in 10 mM Tris-HCl buffer with pH 6.8 was added to 3 µL xanthine oxidase and 50 µL hypoxanthine, in a 96 well plate with a total volume of 100 µL. The superoxide free radicals were generated in the system by hypoxanthine and xanthine oxidase. A sufficient amount of catalase (0.5 µL) was added to the system to avoid the obtrusion of hydrogen peroxide. The kinetics was run for 20 minutes at 550 nm absorbance. The oxidized CNPs was incubated for approximately 15 days followed by SOD activity measurement.

### IV.9 - CAT mimetic activity assay

The catalytic activity of CNPs in the disproportionation of H$_2$O$_2$ was assessed by spectrofluorimetry using the Amplex-Red reagent assay (Cat # A22180).[49-51] Briefly, 25 µL of a nanoparticle dispersion in Tris-Cl buffer pH 7.5 was added to a well of a 96-well plate and mixed with 25 µL of a $H_2O_2$ solution. Then, 50 μL Amplex Red reagent/HRP working solution was added and reactions are pre-incubated for 5 minutes. Amplex Red (10-acetyl-3,7-dihydroxyphenoxazine) reaction with $H_2O_2$ catalyzed by horseradish peroxide (HRP) produces the fluorescent molecule resorufin (excitation at 571 nm and emission at 585 nm). The fluorescence was measured after incubating for 30 min with protection from light. The $H_2O_2$ concentration in each well was 5 μM, whereas those of CNPs ranged from $3.5\times10^{-1}$ to $3.5\times10^{4}$ μM. The catalytic activity noted $A_{CAT}$ is defined as the percentage of decomposed $H_2O_2$ determined at the end of the assay.

### IV.10 – Peroxidase and oxidase mimetic activity assay

The steady-state kinetics of TMB oxidation catalyzed by nanozymes was carried out in a Hitachi UV2010 spectrophotometer by following the absorption of the oxidation product at 652 nm. The experiments were done in acetate buffer 0.1 M pH 4.0 at 25°C, the temperature being controlled by a Peltier. The effect of concentration in the nanozyme peroxidase-like activity was evaluated by varying the nanoparticle molar concentration [CNP] from 1 to 20 nM, maintaining those of TMB and of hydrogen peroxide at 0.8 mM and 2 M, respectively. For the peroxidase-like catalytic activity, kinetic curves were obtained in triplicate using [CNP] = 10 nM, [TMB] = 0 – 0.8 mM and $[H_2O_2]$ = 0 – 2 M. For the oxidase-like catalytic activity, the data were obtained in triplicate by fixing the CNP at 250 nM, with the same conditions as above for [TMB] and $[H_2O_2]$. The kinetic parameters $V_{Max}$ and $K_m$ which are the maximum rate of product formation and the Michaelis-Menten constant were determined after fitting the time dependences with the Michaelis-Menten equation (Eqs. 3 and 4).

### IV.11 – Detection of hydroxyl radical generation

Hydroxyl radical (˙OH) generation was investigated by using terephthalic acid as a probe. A stock solution of 15 mM of terephthalic acid was prepared by adding 124.5 mg in 50 mL of deionized water consisting of 37 mM NaOH. To perform the reaction 5 mM of terephthalic acid was mixed with 10 nM CNPs and 50 mM $H_2O_2$, in a total volume of 3 mL. The reaction mixture was incubated for 30 minutes in dark at room temperature. After that the solution was centrifuged at 10,000 rpm for 5 minutes to remove the background. The fluorescence intensity of conversion of terephthalic acid to 2-Hydroxyterephthalic acid (2HTA) was measured at 315 nm excitation wavelength and emission spectra was followed from 350 to 550 nm using Cary Eclipse fluorescence spectrophotometer (Agilent). 1% ethanol was added to the solution to study the scavenging impact of ethanol on hydroxyl radical generation in the reaction.

### IV.12 – Indirect Quantification of hydroxyl radicals

To quantify the hydroxyl radicals produced in the reaction the fluorescence intensity was compared with the fluorescence of standard 2-Hydroxyterephthalic acid. Various concentrations of 2HTA (2.5 μM, 5 μM, 7.5 μM, 10 μM and 12.5 μM) were prepared in 0.1 X PBS. The standard curve from 2HTA fluorescence was plotted and from that the concentration of hydroxyl radicals generated by cerium oxide nanoparticles in presence and absence of ethanol was calculated and represented.

# Supporting Information

Figure S1: Wide-Angle X-Ray Scattering (WAXS) of cerium oxide nanoparticles – Figure S2 and Table S2 : X-ray photoelectron spectrometry (XPS) Ce 3d spectra of cerium oxide nanoparticles – Figures S3: Copolymers synthesis and characterization using $^1$H NMR – Figures S4: Weight-averaged molecular weight determination of the copolymers obtained from static light scattering – Figure S5: TEM images of polymer coated cerium oxide nanoparticles – Figure S6 and Table S6: Oxidase-like catalytic activity of cerium oxide nanoparticles – Figure S7: Calibration for hydroxyl radical measurements in relation with peroxidase-like activity – Figure S8: Quantification of hydroxyl radicals.

# Acknowledgements

We thank Fethi Bedioui, Virginie Berthat, Jérôme Fresnais, Geoffroy Goujon, Isabelle Margaill, Nathalie Mignet, Evdokia Oikonomou and Caroline Roques for fruitful discussions. Alexandre da Silva is acknowledged for his support for the oxidase and peroxidase experiments. ANR (Agence Nationale de la Recherche) and CGI (Commissariat à l'Investissement d'Avenir) are acknowledged for their financial support of this work through Labex SEAM (Science and Engineering for Advanced Materials and devices) ANR 11 LABX 086. ANR 11 IDEX 05 02. We acknowledge the ImagoSeine facility (Jacques Monod Institute, Paris, France), and the France BioImaging infrastructure supported by the French National Research Agency (ANR-10-INSB-04, « Investments for the future ») for the transmission electron microscopy support. This research was supported in part by the Agence Nationale de la Recherche under the contract ANR-13-BS08-0015 (PANORAMA). ANR-12-CHEX-0011 (PULMONANO) and ANR-15-CE18-0024-01 (ICONS. Innovative polymer coated cerium oxide for stroke treatment).

# TOC image

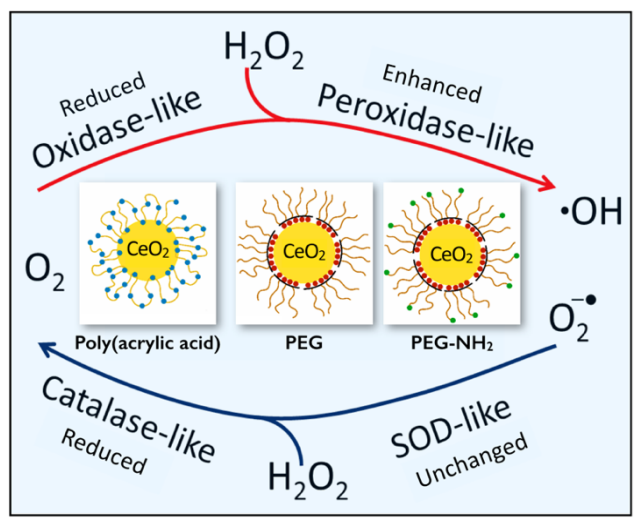

Cerium oxide nanoparticles have demonstrated antioxidant properties, resulting in intensive research into nanomedicine applications. In this study, polymeric coatings using poly (ethylene glycol) and phosphonic groups are found to significantly to improve colloidal stability of nanoparticles in biofluids and also to reduce, preserve or increase the enzyme-like catalytic activities of the cerium oxide.

# References

(1) Campbell, C. T.; Peden, C. H. Oxygen vacancies and catalysis on ceria surfaces. *Science* **2005,** *309* (5735), 713-714.
(2) Esch, F.; Fabris, S.; Zhou, L.; Montini, T.; Africh, C.; Fornasiero, P.; Comelli, G.; Rosei, R. Electron Localization Determines Defect Formation on Ceria Substrates. *Science* **2005,** *309* (5735), 752-755.
(3) Ma, Y.; Gao, W.; Zhang, Z.; Zhang, S.; Tian, Z.; Liu, Y.; Ho, J. C.; Qu, Y. Regulating the surface of nanoceria and its applications in heterogeneous catalysis. *Surf. Sci. Rep.* **2018,** *73* (1), 1-36.
(4) Davis, M. E.; Davis, R. J. *Fundamentals of chemical reaction engineering*, Courier Corporation: 2012.
(5) Trovarelli, A.; Fornasiero, P. *Catalysis by Ceria and Related Materials, 2nd Edition*, Imperial College Press: London, 2013; Vol. 12.
(6) Dong, J.; Song, L.; Yin, J.-J.; He, W.; Wu, Y.; Gu, N.; Zhang, Y. Co3O4 nanoparticles with multi-enzyme activities and their application in immunohistochemical assay. *ACS Appl. Mater. Interfaces* **2014,** *6* (3), 1959-1970.
(7) Gao, L.; Zhuang, J.; Nie, L.; Zhang, J.; Zhang, Y.; Gu, N.; Wang, T.; Feng, J.; Yang, D.; Perrett, S. Intrinsic peroxidase-like activity of ferromagnetic nanoparticles. *Nat. Nanotechnol.* **2007,** *2* (9), 577-583.
(8) He, W.; Cai, J.; Zhang, H.; Zhang, L.; Zhang, X.; Li, J.; Yin, J.-J. Formation of PtCuCo trimetallic nanostructures with enhanced catalytic and enzyme-like activities for biodetection. *ACS Appl. Nano Mater.* **2017,** *1* (1), 222-231.
(9) Liu, B.; Huang, Z.; Liu, J. Boosting the oxidase mimicking activity of nanoceria by fluoride capping: rivaling protein enzymes and ultrasensitive F− detection. *Nanoscale* **2016,** *8* (28), 13562-13567.
(10) Mu, J.; Zhang, L.; Zhao, M.; Wang, Y. Catalase mimic property of Co3O4 nanomaterials with different morphology and its application as a calcium sensor. *ACS Appl. Mater. Interfaces* **2014,** *6* (10), 7090-7098.
(11) Singh, N.; Savanur, M. A.; Srivastava, S.; D'Silva, P.; Mugesh, G. A redox modulatory Mn3O4 nanozyme with multi-enzyme activity provides efficient cytoprotection to human cells in a Parkinson's disease model. *Angew. Chem. Int. Ed.* **2017,** *56* (45), 14267-14271.
(12) Ye, H.; Wang, Q.; Catalano, M.; Lu, N.; Vermeylen, J.; Kim, M. J.; Liu, Y.; Sun, Y.; Xia, X. Ru nanoframes with an fcc structure and enhanced catalytic properties. *Nano Lett.* **2016,** *16* (4), 2812-2817.
(13) Yu, F.; Huang, Y.; Cole, A. J.; Yang, V. C. The artificial peroxidase activity of magnetic iron oxide nanoparticles and its application to glucose detection. *Biomaterials* **2009,** *30* (27), 4716-4722.
(14) Andreescu, D.; Bulbul, G.; Özel, R. E.; Hayat, A.; Sardesai, N.; Andreescu, S. Applications and implications of nanoceria reactivity: measurement tools and environmental impact. *Environ. Sci.: Nano* **2014,** *1* (5), 445-458.
(15) Singh, S. Cerium oxide based nanozymes: Redox phenomenon at biointerfaces. *Biointerphases* **2016,** *11* (4).
(16) Cai, X.; Sezate, S. A.; Seal, S.; McGinnis, J. F. Sustained protection against photoreceptor degeneration in tubby mice by intravitreal injection of nanoceria. *Biomaterials* **2012,** *33* (34), 8771-8781.

(17) Eitan, E.; Hutchison, E. R.; Greig, N. H.; Tweedie, D.; Celik, H.; Ghosh, S.; Fishbein, K. W.; Spencer, R. G.; Sasaki, C. Y.; Ghosh, P. Combination therapy with lenalidomide and nanoceria ameliorates CNS autoimmunity. *Exp. Neurol.* **2015,** *273*, 151-160.

(18) Estevez, A.; Pritchard, S.; Harper, K.; Aston, J.; Lynch, A.; Lucky, J.; Ludington, J.; Chatani, P.; Mosenthal, W.; Leiter, J. Neuroprotective mechanisms of cerium oxide nanoparticles in a mouse hippocampal brain slice model of ischemia. *Free Radical Bio. Med.* **2011,** *51* (6), 1155-1163.

(19) Estevez, A. Y.; Erlichman, J. S. The potential of cerium oxide nanoparticles (nanoceria) for neurodegenerative disease therapy. *Nanomedicine* **2014,** *9* (10), 1437-1440.

(20) Khurana, A.; Tekula, S.; Godugu, C. Nanoceria suppresses multiple low doses of streptozotocin-induced Type 1 diabetes by inhibition of Nrf2/NF-κB pathway and reduction of apoptosis. *Nanomedicine* **2018,** *13* (15), 1905-1922.

(21) Selvaraj, V.; Manne, N. D.; Arvapalli, R.; Rice, K. M.; Nandyala, G.; Fankenhanel, E.; Blough, E. R. Effect of cerium oxide nanoparticles on sepsis induced mortality and NF-κB signaling in cultured macrophages. *Nanomedicine* **2015,** *10* (8), 1275-1288.

(22) Selvaraj, V.; Nepal, N.; Rogers, S.; Manne, N. D.; Arvapalli, R.; Rice, K. M.; Asano, S.; Fankhanel, E.; Ma, J. J.; Shokuhfar, T. Inhibition of MAP kinase/NF-kB mediated signaling and attenuation of lipopolysaccharide induced severe sepsis by cerium oxide nanoparticles. *Biomaterials* **2015,** *59*, 160-171.

(23) Walkey, C.; Das, S.; Seal, S.; Erlichman, J.; Heckman, K.; Ghibelli, L.; Traversa, E.; McGinnis, J. F.; Self, W. T. Catalytic properties and biomedical applications of cerium oxide nanoparticles. *Environ. Sci.: Nano* **2015,** *2* (1), 33-53.

(24) Wei, H.; Wang, E. Nanomaterials with enzyme-like characteristics (nanozymes): next-generation artificial enzymes. *Chem. Soc. Rev.* **2013,** *42* (14), 6060-6093.

(25) Asati, A.; Santra, S.; Kaittanis, C.; Nath, S.; Perez, J. M. Oxidase-Like Activity of Polymer-Coated Cerium Oxide Nanoparticles. *Angew. Chem. Int. Ed.* **2009,** *48* (13), 2308-2312.

(26) Patel, V.; Singh, M.; Mayes, E. L. H.; Martinez, A.; Shutthanandan, V.; Bansal, V.; Singh, S.; Karakoti, A. S. Ligand-mediated reversal of the oxidation state dependent ROS scavenging and enzyme mimicking activity of ceria nanoparticles. *Chem. Commun.* **2018,** *54* (99), 13973-13976.

(27) Baldim, V.; Bia, N.; Graillot, A.; Loubat, C.; Berret, J.-F. Monophosphonic versus Multiphosphonic Acid Based PEGylated Polymers for Functionalization and Stabilization of Metal (Ce, Fe, Ti, Al) Oxide Nanoparticles in Biological Media. *Adv. Mater. Interfaces* **2019,** *6* (7), 1801814.

(28) Ould-Moussa, N.; Safi, M.; Guedeau-Boudeville, M.-A.; Montero, D.; Conjeaud, H.; Berret, J.-F. In vitro toxicity of nanoceria: effect of coating and stability in biofluids. *Nanotoxicology* **2014,** *8* (7), 799-811.

(29) Torrisi, V.; Graillot, A.; Vitorazi, L.; Crouzet, Q.; Marletta, G.; Loubat, C.; Berret, J.-F. Preventing Corona Effects: Multiphosphonic Acid Poly(ethylene glycol) Copolymers for Stable Stealth Iron Oxide Nanoparticles. *Biomacromolecules* **2014,** *15* (8), 3171-3179.

(30) Berret, J.-F.; Sehgal, A.; Morvan, M.; Sandre, O.; Vacher, A.; Airiau, M. Stable oxide nanoparticle clusters obtained by complexation. *J. Colloid Interface Sci.* **2006,** *303* (1), 315-318.

(31) Qi, L.; Chapel, J. P.; Castaing, J. C.; Fresnais, J.; Berret, J.-F. Organic versus hybrid coacervate complexes: co-assembly and adsorption properties. *Soft Matter* **2008,** *4* (3), 577-585.

(32) Baldim, V.; Bedioui, F.; Mignet, N.; Margaill, I.; Berret, J. F. The enzyme-like catalytic activity of cerium oxide nanoparticles and its dependency on Ce3+ surface area concentration. *Nanoscale* **2018,** *10* (15), 6971-6980.

(33) Qi, L.; Sehgal, A.; Castaing, J.-C.; Chapel, J.-P.; Fresnais, J.; Berret, J.-F.; Cousin, F. Redispersible Hybrid Nanopowders: Cerium Oxide Nanoparticle Complexes with Phosphonated-PEG Oligomers. *ACS Nano* **2008,** *2* (5), 879-888.

(34) Das, M.; Mishra, D.; Dhak, P.; Gupta, S.; Maiti, T. K.; Basak, A.; Pramanik, P. Biofunctionalized, Phosphonate-Grafted, Ultrasmall Iron Oxide Nanoparticles for Combined Targeted Cancer Therapy and Multimodal Imaging. *Small* **2009,** *5* (24), 2883-2893.

(35) Giamblanco, N.; Marletta, G.; Graillot, A.; Bia, N.; Loubat, C.; Berret, J.-F. Serum Protein-Resistant Behavior of Multisite-Bound Poly(ethylene glycol) Chains on Iron Oxide Surfaces. *ACS Omega* **2017,** *2* (4), 1309-1320.

(36) Graillot, A.; Monge, S.; Faur, C.; Bouyer, D.; Robin, J.-J. Synthesis by RAFT of innovative well-defined (co) polymers from a novel phosphorus-based acrylamide monomer. *Polym. Chem.* **2013,** *4* (3), 795-803.

(37) Sandiford, L.; Phinikaridou, A.; Protti, A.; Meszaros, L. K.; Cui, X. J.; Yan, Y.; Frodsham, G.; Williamson, P. A.; Gaddum, N.; Botnar, R. M.; Blower, P. J.; Green, M. A.; de Rosales, R. T. M. Bisphosphonate-Anchored PEGylation and Radiolabeling of Superparamagnetic Iron Oxide. *ACS Nano* **2013,** *7* (1), 500-512.

(38) Zoulalian, V.; Zurcher, S.; Tosatti, S.; Textor, M.; Monge, S.; Robin, J. J. Self-Assembly of Poly(ethylene glycol)-Poly(alkyl phosphonate) Terpolymers on Titanium Oxide Surfaces: Synthesis, Interface Characterization, Investigation of Nonfouling Properties, and Long-Term Stability. *Langmuir* **2010,** *26* (1), 74-82.

(39) Chanteau, B.; Fresnais, J.; Berret, J.-F. Electrosteric Enhanced Stability of Functional Sub-10 nm Cerium and Iron Oxide Particles in Cell Culture Medium. *Langmuir* **2009,** *25* (16), 9064-9070.

(40) Ramniceanu, G.; Doan, B. T.; Vezignol, C.; Graillot, A.; Loubat, C.; Mignet, N.; Berret, J. F. Delayed hepatic uptake of multi-phosphonic acid poly(ethylene glycol) coated iron oxide measured by real-time magnetic resonance imaging. *RSC Adv.* **2016,** *6* (68), 63788-63800.

(41) Sehgal, A.; Lalatonne, Y.; Berret, J.-F.; Morvan, M. Precipitation-redispersion of cerium oxide nanoparticles with poly(acrylic acid): Toward stable dispersions. *Langmuir* **2005,** *21* (20), 9359-9364.

(42) Karakoti, A. S.; Yang, P.; Wang, W.; Patel, V.; Martinez, A.; Shutthanandan, V.; Seal, S.; Thevuthasan, S. Investigation of the Ligand–Nanoparticle Interface: A Cryogenic Approach for Preserving Surface Chemistry. *J. Phys. Chem. C* **2018,** *122* (6), 3582-3590.

(43) Sanghavi, S.; Wang, W.; Nandasiri, M. I.; Karakoti, A. S.; Wang, W.; Yang, P.; Thevuthasan, S. Investigation of trimethylacetic acid adsorption on stoichiometric and oxygen-deficient CeO2(111) surfaces. *Phys. Chem. Chem. Phys.* **2016,** *18* (23), 15625-15631.

(44) Guerrero, G.; Mutin, P. H.; Vioux, A. Anchoring of Phosphonate and Phosphinate Coupling Molecules on Titania Particles. *Chem. Mater.* **2001,** *13* (11), 4367-4373.

(45) Paniagua, S. A.; Giordano, A. J.; Smith, O. N. L.; Barlow, S.; Li, H.; Armstrong, N. R.; Pemberton, J. E.; Bredas, J.-L.; Ginger, D.; Marder, S. R. Phosphonic Acids for Interfacial Engineering of Transparent Conductive Oxides. *Chem. Rev.* **2016,** *116* (12), 7117-7158.

(46) Pawsey, S.; Yach, K.; Reven, L. Self-Assembly of Carboxyalkylphosphonic Acids on Metal Oxide Powders. *Langmuir* **2002,** *18* (13), 5205-5212.

(47) Brittain, W. J.; Minko, S. A structural definition of polymer brushes. *J. Polym. Sci. Pol. Chem.* **2007,** *45* (16), 3505-3512.

(48) Milner, S. T. Polymer brushes. *Science* **1991,** *251* (4996), 905-914.

(49) Gil, D.; Rodriguez, J.; Ward, B.; Vertegel, A.; Ivanov, V.; Reukov, V. Antioxidant Activity of SOD and Catalase Conjugated with Nanocrystalline Ceria. *Bioeng.* **2017,** *4* (1), 18-18.

(50) Kim, C. K.; Kim, T.; Choi, I.-Y.; Soh, M.; Kim, D.; Kim, Y.-J.; Jang, H.; Yang, H.-S.; Kim, J. Y.; Park, H.-K.; Park, S. P.; Park, S.; Yu, T.; Yoon, B.-W.; Lee, S.-H.; Hyeon, T. Ceria Nanoparticles that can Protect against Ischemic Stroke. *Angew. Chemie Int. Ed.* **2012,** *51* (44), 11039-11043.

(51) Soh, M.; Kang, D.-w.; Jeong, H.-g.; Kim, D.; Kim, D. Y.; Yang, W.; Song, C.; Baik, S.; Choi, I.-y.; Ki, S.-k.; Kwon, H. J.; Kim, T.; Kim, C. K.; Lee, S.-h.; Hyeon, T. Ceria-Zirconia Nanoparticles as an Enhanced Multi-Antioxidant for Sepsis Treatment. *Angew. Chemie Int. Ed.* **2017**, *56*, 11399-11403

(52) Lee, S. S.; Song, W.; Cho, M.; Puppala, H. L.; Nguyen, P.; Zhu, H.; Segatori, L.; Colvin, V. L. Antioxidant Properties of Cerium Oxide Nanocrystals as a Function of Nanocrystal Diameter and Surface Coating. *ACS Nano* **2013**, *7* (11), 9693-9703.

(53) Ornatska, M.; Sharpe, E.; Andreescu, D.; Andreescu, S. Paper Bioassay Based on Ceria Nanoparticles as Colorimetric Probes. *Anal. Chem.* **2011**, *83* (11), 4273-4280.

(54) Damatov, D.; Mayer, J. M. (Hydro)peroxide ligands on colloidal cerium oxide nanoparticles. *Chem. Commun.* **2016**, *52* (52), 10281-10284.

(55) Scholes, F. H.; Soste, C.; Hughes, A. E.; Hardin, S. G.; Curtis, P. R. The role of hydrogen peroxide in the deposition of cerium-based conversion coatings. *Appl. Surf. Sci.* **2006**, *253* (4), 1770-1780.

(56) Wang, Y.-J.; Dong, H.; Lyu, G.-M.; Zhang, H.-Y.; Ke, J.; Kang, L.-Q.; Teng, J.-L.; Sun, L.-D.; Si, R.; Zhang, J.; Liu, Y.-J.; Zhang, Y.-W.; Huang, Y.-H.; Yan, C.-H. Engineering the defect state and reducibility of ceria based nanoparticles for improved anti-oxidation performance. *Nanoscale* **2015**, *7* (33), 13981-13990.

(57) Zang, C.; Zhang, X.; Hu, S.; Chen, F. The role of exposed facets in the Fenton-like reactivity of CeO$_2$ nanocrystal to the Orange II. *Appl. Catal. B Environ.* **2017**, *216*, 106-113.

(58) Tian, Z.; Li, X.; Ma, Y.; Chen, T.; Xu, D.; Wang, B.; Qu, Y.; Gao, Y. Quantitatively Intrinsic Biomimetic Catalytic Activity of Nanocerias as Radical Scavengers and Their Ability against H$_2$O$_2$ and Doxorubicin-Induced Oxidative Stress. *ACS Appl. Mater. Interfaces* **2017**, *9* (28), 23342-23352.

(59) Karakoti, A.; Singh, S.; Dowding, J. M.; Seal, S.; Self, W. T. Redox-active radical scavenging nanomaterials. *Chem. Soc. Rev.* **2010**, *39* (11), 4422-4422.

(60) Veitch, N. C. Horseradish peroxidase: a modern view of a classic enzyme. *Phytochemistry* **2004**, *65* (3), 249-259.

(61) Shah, J.; Purohit, R.; Singh, R.; Karakoti, A. S.; Singh, S. ATP-enhanced peroxidase-like activity of gold nanoparticles. *J. Colloid Interface Sci.* **2015**, *456*, 100-107.

(62) Vallabani, N. V. S.; Karakoti, A. S.; Singh, S. ATP-mediated intrinsic peroxidase-like activity of Fe3O4-based nanozyme: One step detection of blood glucose at physiological pH. *Colloids Surf., B* **2017**, *153*, 52-60.

(63) Vallabani, N. V. S.; Singh, S.; Karakoti, A. S. Investigating the role of ATP towards amplified peroxidase activity of Iron oxide nanoparticles in different biologically relevant buffers. *Appl. Surf. Sci.* **2019**, *492*, 337-348.

(64) Gonzalez, D. H.; Kuang, X. M.; Scott, J. A.; Rocha, G. O.; Paulson, S. E. Terephthalate probe for hydroxyl radicals: Yield of 2-hydroxyterephthalic acid and transition metal interference. *Anal. Lett.* **2018**, *51* (15), 2488-2497.

(65) Fresnais, J.; Yan, M.; Courtois, J.; Bostelmann, T.; Bee, A.; Berret, J. F. Poly(acrylic acid)-coated iron oxide nanoparticles: Quantitative evaluation of the coating properties and applications for the removal of a pollutant dye. *J. Colloid Interface Sci.* **2013**, *395*, 24-30.

(66) Dhont, J. K. G. *An Introduction to Dynamics of Colloids*, Elsevier: Amsterdam, 1996.

(67) Korsvik, C.; Patil, S.; Seal, S.; Self, W. T. Superoxide dismutase mimetic properties exhibited by vacancy engineered ceria nanoparticles. *Chem. Commun.* **2007**, (10), 1056-1056.